\documentclass[aps,prl,twocolumn,superscriptaddress,
  preprintnumbers,nofootinbib]{revtex4-1}
\usepackage[hyperindex,breaklinks,colorlinks=true,citecolor=blue,hyperfootnotes=false]{hyperref}
\usepackage{amsmath}
\usepackage{amsfonts}
\usepackage{amssymb}
\usepackage{mathrsfs}
\usepackage{graphicx}
\usepackage{xcolor}

\usepackage{soul}

\DeclareMathOperator{\arctanh}{\mathrm{arctanh}}
\DeclareMathOperator{\Li}{\mathrm{Li}}

\usepackage{tikz}
\usetikzlibrary{decorations.pathmorphing,calc}

\tikzset{
  wilson/.style={line width=2pt},
  gluon/.style={
    decorate,
    decoration={coil,aspect=0.55,segment length=3.8pt,amplitude=1.7pt},
    line width=0.8pt
  },
  selfenergy/.style={
    draw,
    circle,
    minimum size=9mm,
    fill=gray!12,
    line width=0.8pt
  },
  lab/.style={font=\small}
}

\begin{document}
\preprint{MIT-CTP/6020}

\title{Heavy Quark Transport is Non-Gaussian Beyond Leading Log}

\author{Jean F. Du Plessis}
\email[]{jeandp@mit.edu}
\affiliation{Center for Theoretical Physics --- A Leinweber Institute,\\ Massachusetts Institute of Technology, Cambridge, Massachusetts 02139, USA}
\affiliation{Kavli Institute for Theoretical Physics,\\ University of California, Santa Barbara, California 93106, USA}

\author{Bruno Scheihing-Hitschfeld}
\email[]{bscheihi@kitp.ucsb.edu}
\affiliation{Kavli Institute for Theoretical Physics,\\ University of California, Santa Barbara, California 93106, USA}

\date{\today}

\begin{abstract}
We find that heavy quark transport beyond leading logarithm at weak coupling is intrinsically non-Gaussian: the longitudinal momentum transfer distribution has \textit{asymmetric exponential tails} that are crucial for equilibration dynamics. We show this by computing the leading-order momentum transfer kernel for relativistic heavy quarks in weakly coupled non-Abelian plasmas, matching perturbative momentum transfer on the thermal scale to hard-thermal-loop-resummed soft physics. This is the same structure previously found in strongly coupled holographic plasmas, showing that it is not peculiar to weak or strong coupling, conformality, or supersymmetry. We therefore expect that this is a robust feature that physical quark-gluon plasma should also exhibit.

\end{abstract}

\maketitle

Heavy quarks (HQs) provide a particularly valuable probe of the quark-gluon plasma (QGP) produced in ultra-relativistic heavy-ion collisions (HICs).
Because they are long-lived on quantum chromodynamics (QCD) time scales and interact directly with the thermal medium, their propagation offers a sensitive window into how colored probes exchange energy and momentum with QGP constituents. This separation of scales makes heavy quarks natural candidates for a Brownian-motion-like description, and motivated early Langevin and Fokker--Planck treatments of heavy-flavor transport~\cite{Moore:2004tg,vanHees:2004gq,vanHees:2005wb,Mustafa:2004dr} shortly after data became available. Together with Boltzmann transport approaches, they constitute the state of the art of heavy quark transport~\cite{Prino:2016cni,Xu:2018gux,Rapp:2018qla,Cao:2018ews,Dong:2019unq,He:2022ywp}. The non-perturbative determination of the HQ diffusion coefficient~\cite{Moore:2004tg,Casalderrey-Solana:2006fio}, the essential QCD input for a Langevin description, is by now mature in the $v\to0$ limit using lattice QCD methods~\cite{Brambilla:2020siz,Altenkort:2020fgs,Brambilla:2022xbd,Altenkort:2023oms,Altenkort:2023eav,HotQCD:2025fbd} in tandem with perturbative calculations~\cite{Caron-Huot:2007rwy,Caron-Huot:2008dyw,Burnier:2010rp,Scheihing-Hitschfeld:2023tuz,delaCruz:2024cix}, providing an increasingly sharp first-principles input for phenomenological comparisons to heavy-ion collision data.

However, challenges still remain.
It has long been understood \cite{Moore:2004tg} that, for a heavy quark moving through a weakly coupled plasma, the Einstein relation $\kappa_L = 2 M T \gamma \eta_D$ between the drag $\eta_D$ and longitudinal diffusion coefficients $\kappa_L$ --- the necessary condition for kinetic equilibration to take place in a (Gaussian) Langevin description --- is violated at nonzero velocity once one goes beyond leading logarithmic accuracy. The Einstein relation is also violated in strongly coupled $\mathcal{N}=4$ supersymmetric Yang–Mills (SYM) theory~\cite{Herzog:2006gh,Gubser:2006bz,Gubser:2006nz,Casalderrey-Solana:2007ahi}. In a purely Langevin/Fokker--Planck description this is a serious obstruction to connecting first-principles calculations with phenomenological descriptions~\cite{Peshier:2008bg}. Until now, this issue had not been resolved in perturbative QCD. Furthermore, the problem is not merely formal. Boltzmann transport studies have shown that non-Gaussian momentum transfer dynamics can modify heavy-flavor observables and the transport coefficients extracted from them~\cite{Das:2013kea,Scardina:2017ipo}, which may be relevant to the tension between phenomenological extractions and lattice QCD determinations of heavy quark transport coefficients~\cite{Beraudo:2025nvq}. 

Recent work has clarified the conceptual resolution of this issue. When retaining the full momentum transfer statistics of the process, heavy quark evolution is governed not by a Fokker--Planck equation but by a Kolmogorov equation, and equilibration is controlled by a universal non-Gaussian condition on the corresponding kernel~\cite{Rajagopal:2025rxr}. In strongly coupled $\mathcal N=4$ SYM, the full leading order kernel was shown to satisfy this generalized equilibration condition and to exhibit a distinctive non-Gaussian longitudinal momentum transfer distribution~\cite{Rajagopal:2025ukd}.
This raises a natural question: is the same non-Gaussian structure present for heavy quarks in weakly coupled gauge theory plasmas?

In this Letter we answer that question in the affirmative. We find that heavy quark transport at weak coupling beyond leading logarithm exhibits manifest non-Gaussian structure: the longitudinal momentum transfer distribution has a Gaussian core and asymmetric exponential tails. The tails are essential for equilibration in light of the violation of the Einstein relation~\cite{Rajagopal:2025rxr}, meaning that HQ transport is intrinsically non-Gaussian. The Gaussian core follows as a natural consequence of the central limit theorem. On the other hand, the exponential tails are a new and surprising result. In fact, this structure is also present in the strongly coupled calculation of~\cite{Rajagopal:2025ukd} --- a fact that has remained underappreciated so far.
With weakly coupled QCD and both weakly \textit{and} strongly coupled $\mathcal{N}=4$ SYM exhibiting this same structure, this indicates that Gaussian core and exponential tail momentum transfer is not peculiar to weak or strong coupling, conformality, or supersymmetry, but is instead likely to be a robust feature of realistic QGP. To establish this result, we consistently match the ultraviolet (UV) perturbative contribution to the infrared hard thermal loop (HTL) sector, isolate the strict fixed-order kernel, and determine the longitudinal momentum transfer distribution at all momenta via a Legendre transform. The full matched kernel reproduces the known leading-log results for the drag and momentum diffusion coefficients, as well as the known leading-order diffusion coefficient at ${\bf v}={\bf 0}$, thereby recovering the appropriate Gaussian limits. It also satisfies the generalized equilibration condition of Ref.~\cite{Rajagopal:2025rxr}, providing a nontrivial consistency check on the resulting non-Gaussian dynamics.

\textbf{Setup.}
We work in the large mass limit, neglecting all mass-dependent effects at the level of transition probabilities. Formally, we work at leading order in the heavy quark effective theory (HQET) expansion of the QCD Lagrangian~\cite{Georgi:1990um}, neglecting $1/M$ corrections. In this regime the momentum transfer distribution of the heavy quark to the medium over a time $\Delta t$ is determined by the Fourier transform of a Wilson loop
\begin{equation}
    P({\bf k};{\bf v})=\int \frac{d^3{\bf L}}{(2\pi)^3}\,
    e^{-i{\bf k}\cdot{\bf L}}\,
    \langle W_{\bf v}\rangle({\bf L};\Delta t),
    \label{eq:P-from-W}
\end{equation}
which can be formulated on the Schwinger--Keldysh contour~\cite{Rajagopal:2025ukd,Rajagopal:2025rxr}. In the late-time limit, the separation of scales $M\gg T$ implies that heavy quarks undergo Markovian dynamics due to the shortness of the medium equilibration time scale $\sim 1/T$ relative to the time needed to appreciably modify the velocity of the heavy quark. For such time scales $\Delta t \gg 1/T$, we have
\begin{equation}
    \langle W_{\bf v}\rangle({\bf L};\Delta t)=
    \exp\!\left[-(\Delta t)\,T\,S({\bf L};{\bf v})+\ldots\right],
    \label{eq:W-S}
\end{equation}
which defines the momentum transfer kernel $S({\bf L};{\bf v})$.

At weak coupling, a consistent leading-order result requires matching perturbative ``hard'' momentum transfer $(p\sim T)$ (following from finite temperature Feynman diagrams~\cite{Keldysh:1964ud,Schwinger:1960qe,Laine:2016hma,Ghiglieri:2020dpq}) to HTL ``soft'' physics~\cite{Braaten:1989mz,Braaten:1990az,Braaten:1991gm,Caron-Huot:2007cma} $(p\sim gT)$ while subtracting their overlap,
\begin{equation}
    ({\rm Result})=({\rm Pert.})+({\rm HTL})-({\rm HTL,\ unresummed}) \, .
    \label{eq:matching-schematic}
\end{equation}
This procedure has been used to calculate the HQ diffusion coefficient~\cite{Burnier:2010rp}.
By definition, sub-extensive contributions in $\Delta t$ do not contribute to $S$, and by unitarity of the Wilson lines that make up the loop we have
\begin{equation}
    S({\bf L}={\bf 0};{\bf v})=0 \, .
\end{equation}
Consequently, the calculation reduces to extracting only those contributions that are simultaneously ${\bf L}$ dependent and extensive in $\Delta t$.
Denoting $\int_{\mathbf{p}} \equiv \int d^3\mathbf{p}/(2\pi)^3$, the matched weak-coupling kernel may then be written as
\begin{widetext}
\begin{equation}
    S({\bf L};{\bf v})
    =
    -\frac{g^2 C_F}{T} \int_{\bf p}
    \frac{e^{i{\bf p}\cdot{\bf L}}-1}{1-e^{-{\bf p}\cdot{\bf v}/T}}
    \left\{g^2
    \mathcal  I_{\rm Hard}({\bf p},{\bf v};T,N_c,N_X,N_f)
    +
    \mathcal I_{\rm Soft}({\bf p},{\bf v};m_D^2)
    -
    \mathcal I_{\rm Match}({\bf p},{\bf v};m_D^2)
    \right\},
    \label{eq:kernel}
\end{equation}
\end{widetext}
where we give the full form of each term in the End Matter. We also note the appearance of the Debye mass  $m_D^2 = g^2 T^2[N_c/3 + N_c N_X/6 + T_F N_f/3]$ in the soft and matching integrands\footnote{We denote the number of ``colors'' of the SU$(N_c)$ gauge group by $N_c$, the number of real adjoint scalars in the theory by $N_X$, and the number of Dirac fermions in representation $F$ by $N_f$. $T_F$ is the index of the representation ${\rm Tr}[T^a_F T^b_F] = T_F \delta^{ab}$.}. Its structure immediately implies
\begin{equation}
    S({\bf L};{\bf v})
    =
    S(i{\bf v}/T-{\bf L};{\bf v}),
    \label{eq:gen-eq}
\end{equation}
which is precisely the generalized equilibration condition derived in Ref.~\cite{Rajagopal:2025rxr}. Its small velocity expansion recovers the Einstein relation, whereby a Gaussian truncation becomes consistent with equilibration as $v \to 0$, although higher (even) cumulants generically remain nonzero at $v=0$. Einstein-relation violation is therefore a finite velocity effect, but non-Gaussianity is present at all velocities. For present purposes, however, the central role of Eq.~\eqref{eq:kernel} is to define the longitudinal momentum-transfer distribution whose shape we analyze below in a controlled weak-coupling expansion. The symmetry in Eq.~\eqref{eq:gen-eq} then provides a nontrivial consistency check on that construction. Because this symmetry is a structural property of the integrand in Eq.~\eqref{eq:kernel}, independent of the specific forms of $\mathcal{I}_{\rm Hard}$, $\mathcal{I}_{\rm Soft}$, and $\mathcal{I}_{\rm Match}$, the equilibration condition holds order by order in perturbation theory.

\textbf{Fixed-order longitudinal kernel.}
To determine the shape of the weak-coupling momentum transfer distribution beyond leading logarithm, we must first isolate the strict fixed-order kernel. The matched kernel in Eq.~\eqref{eq:kernel} is correct through $\mathcal O(g^4)$, but it contains contamination from $\mathcal O(g^6)$ and higher terms generated by the HTL resummation.

This contamination becomes especially problematic when one attempts to extract information about the distribution from derivatives of the kernel at ${\bf L}=0$. The relevant quantities are the cumulants of the momentum transfer probability per unit time and temperature,
\begin{equation}
    M_{i_1 \ldots \, i_n}({\bf v}) = -
    \left.
    \frac{\partial^n S({\bf L};{\bf v})}
    {\partial(iL_{i_1})\cdots \partial(iL_{i_n})}
    \right|_{{\bf L}=0}  \, , \label{eq:cumulants}
\end{equation}
which characterize the statistical properties of $P({\bf k};{\bf v})$, including quantities of high phenomenological interest such as the drag coefficient $\eta_D$ encoded in $M_i$ and the momentum broadening coefficients $\kappa_T,\kappa_L$ encoded in $M_{ij}$. 
 However, the cumulants in Eq.~\eqref{eq:cumulants} 
become UV divergent for $n\geq 3$ due to the HTL contributions in Eq.~\eqref{eq:kernel}.
This obstructs the determination of the fixed-order kernel, and is also clearly unphysical. This is because for $|{\bf p}| \gg g T$ one has\footnote{By construction, $\mathcal{I}_{\rm Match}$ is the leading term in the expansion of $\mathcal{I}_{\rm Soft}$ in powers of $m_D^2/|{\bf p}|^2$, since the HTL-resummed propagator depends on the coupling and temperature only through $m_D^2$. Because $\mathcal{I}_{\rm Match} \propto m_D^2/|{\bf p}|^4$, the difference~\eqref{eq:high-p-HTL-cancel} follows as written. }
\begin{equation}
    \mathcal I_{\rm Soft}({\bf p},{\bf v};m_D^2)-\mathcal I_{\rm Match}({\bf p},{\bf v};m_D^2)
    =
    \mathcal O\!\left(\frac{m_D^4}{|{\bf p}|^6}\right) \, , \label{eq:high-p-HTL-cancel}
\end{equation}
which, together with the overall $g^2$ prefactor in~\eqref{eq:kernel}, is a strictly higher order contribution than the $g^4$ terms we are aiming to determine, and nonetheless yields an infinite contribution to the cumulants because of the aforementioned UV divergence. 

A resolution to this problem can be achieved by inserting Heaviside step functions to separate the contributions in each regime (reminiscent of the original methods by Braaten and Thoma~\cite{Braaten:1991jj,Braaten:1991we}). Concretely, writing
\begin{widetext}
    \begin{align}
    S({\bf L};{\bf v}) = - \frac{g^2 C_F}{T} \int_{\bf p}
    &\frac{e^{i{\bf p}\cdot{\bf L}}-1}{1-e^{-{\bf p}\cdot{\bf v}/T}}
    \big[\theta(|{\bf p}|-T)+\theta(T-|{\bf p}|)\big]
    \left\{ g^2 \mathcal I_{\rm Hard}+\mathcal I_{\rm Soft}-\mathcal I_{\rm Match}\right\} \, ,
    \label{eq:split}
\end{align}
\end{widetext}
we may simply omit the HTL pieces for $|{\bf p}| > T$, since $\mathcal I_{\rm Soft}-\mathcal I_{\rm Match}=\mathcal O\!\left(\frac{m_D^4}{p^6}\right)$. At this stage the result is manifestly fixed-order except for the HTL-resummed contribution for $|{\bf p}| < T$.

In this (soft) region the $g\to 0$ limit and the momentum integral do not commute, so one must first perform the radial integral exactly and only then expand in $m_D/T \propto g$ in order to correctly determine the logarithmically enhanced $g^4 \ln (1/g)$ terms.
The result is
\begin{widetext}
    \begin{align}
    S_{\rm FO}({\bf L};{\bf v}) = &\frac{g^2 C_F m_D^2}{4\pi}\ln\!\left(\frac{m_D}{T}\right)
    \Bigg[
    \frac{(v^2-1)\arctanh(v)+v}{2v^3}\,
    iL_\parallel\!\left(iL_\parallel+\frac{v}{T}\right)+\frac{3v^3+(1-v^2)^2\arctanh(v)-v}{4v^3}(iL_\perp)^2
    \Bigg]
    \nonumber\\
    +&\frac{g^2 C_F m_D^2}{4\pi}
    \int_0^1 d\mu
    \Bigg[
    \mu^2\,iL_\parallel\!\left(iL_\parallel+\frac{v}{T}\right)
    +\frac{1-\mu^2}{2}(iL_\perp)^2
    \Bigg]\Bigg[
    \Phi\!\left(a_E,b_E\right)
    +\frac{v^2(1-\mu^2)}{2(1-v^2\mu^2)}
    \Phi\!\left(a_T,b_T\right)
    \Bigg] \nonumber \\
    - &\frac{g^2 C_F}{T} \int_{\bf p}
    \left\{ g^2
    \frac{e^{i{\bf p}\cdot{\bf L}}-1}{1-e^{-{\bf p}\cdot{\bf v}/T}}\, \mathcal I_{\rm Hard}
    -\theta(T-|{\bf p}|)\, ({\bf p} \cdot i{\bf L})  \left[{\bf p} \cdot\left(i{\bf L} +\frac{{\bf v}}{T} \right)\right] \frac{T  }{2\,{\bf p}\! \cdot\!{\bf v}} \, \mathcal{I}_{\rm Match}
    \right\},
    \label{eq:SFO-main}
\end{align}
\end{widetext}
where
\begin{equation}
    \Phi(a,b)\equiv
    \frac14\ln(a^2+b^2)+\frac{a}{2b}\arctan\!\left(\frac{b}{a}\right),
\end{equation}
and the functions $a_{E,T}$ and $b_{E,T}$ are defined in terms of $\eta=\mu v$ as
\begin{align}
    a_E\equiv 1-\frac{\eta}{2}\ln\!\left(\frac{1+\eta}{1-\eta}\right) \, , & & 
    b_E\equiv \frac{\pi}{2}\eta \, , \\
    a_T\equiv \frac{1}{2}\left[\frac{\eta^2}{1-\eta^2}+\frac{\eta}{2}\ln\!\left(\frac{1+\eta}{1-\eta}\right)\right] \, , & &
    b_T\equiv \frac{\pi}{4}\eta \, .
\end{align}

Eq.~\eqref{eq:SFO-main} is the complete fixed-order (denoted by the subscript ``FO") kernel at order $g^4$.  Its structure cleanly separates two physically distinct pieces. The first line of Eq.~\eqref{eq:SFO-main} contains the logarithmically enhanced piece\footnote{It is also reassuring to note that the longitudinal part of the HTL-resummed contributions (in the first two lines) in the kernel~\eqref{eq:SFO-main} satisfies the Einstein relation. This is mandated by the fact that in the extreme weakly coupled limit equilibration should still be attained --- and the fact that this part of the kernel is quadratic means that the associated stochastic process is Gaussian.}, which we could also have obtained by directly translating the results of~\cite{Moore:2004tg} for the transport coefficients $\eta_D, \kappa_T, \kappa_L$ into our notation, and thus serves as a cross-check of our result. The last two lines contain the remaining fixed-order contribution, determined by the perturbative piece $\mathcal{I}_{\rm Hard}$ and the remaining $g^4$ terms coming from the HTL-resummed result and the matching. In particular, the last line generates arbitrarily high-order cumulants, all of which are finite. This can be seen by examining the exponential suppression of $\mathcal{I}_{\rm Hard}({\bf p},{\bf v})$ at large ${\bf p}$ by statistical factors (see End Matter).

We have explicitly verified that replacing the separation scale $T$ by any $\Lambda\sim T$ leaves the fixed-order kernel $S_{\rm FO}$ invariant. 
As an additional cross-check, we have verified that the second moment of the kernel at ${\bf v}={\bf 0}$ reproduces the known diffusion coefficient~\cite{Moore:2004tg,Burnier:2010rp} exactly at order $g^4$. We discuss both in the End Matter; further derivations are provided in the Supplemental Material.

\textbf{Momentum transfer probability.}
To examine the physical implications of the fixed-order kernel, we proceed along the lines of~\cite{Rajagopal:2025rxr} to calculate the momentum transfer probability in the late-time limit. For the equilibration dynamics of an isotropic distribution function, only the longitudinal projection contributes. We therefore start by introducing the longitudinal evolution kernel
\begin{equation}
    \!\!\!K(x,v)\!\equiv\!
    S_{\rm FO}({\bf L}\!=\!-ix\hat{\bf v};{\bf v}\!=\!v\hat{\bf v})
    \!=
    \!-\!\sum_{n=1}^\infty \frac{\mathfrak{M}_n(v)}{n!}x^n,
    \label{eq:Kxv}
\end{equation}
where $\mathfrak{M}_n(v) \equiv \hat{\bf v}_{i_1}\! \cdots \hat{\bf v}_{i_n} M^{\rm FO}_{i_1 \ldots i_n}$ are the projected (fixed order) cumulants per unit time of the momentum change probability distribution obtained from the Taylor series of $S_{FO}$ as in Eq.~\eqref{eq:cumulants}.

Inside the interval $-v/T < x < 0 $, $K(x,v)$ is directly evaluated from the integral representation in Eq.~\eqref{eq:SFO-main}. Outside this region, because all of its derivatives are finite and calculable, for $x >0$ we determine it using the Taylor expansion around $x=0$, and with the corresponding expansion around $x=-v/T$ implied by Eq.~\eqref{eq:gen-eq} for $x < -v/T$. The agreement of these expansions with the integral representation in their common domain provides a nontrivial check that the analytic continuation is under control.

In the Markovian limit, following~\cite{Rajagopal:2025rxr}, the longitudinal momentum transfer probability takes the form
\begin{equation}
    P(k_L;v) = \exp \left( - t T \tilde{S}( C_L ; v ) + \ldots \right) \, , \label{eq:P-H}
\end{equation}
where $C_L = {\bf k} \cdot \hat{\bf v} /(t T)$ is the longitudinal momentum transfer rate, the dots ``$\ldots$'' stand for non-extensive terms in $t$, 
and $\tilde{S}$
is the Legendre transform of the evolution kernel $K(x,v)$
\begin{equation}
    \tilde{S}(C_L;v) = \left[ K(x,v) - x \frac{\partial K}{\partial x} \right]_{{x} = {x}({C_L})} \, , \label{eq:Legendre-tf}
\end{equation}
where $x(C_L)$ is determined by the implicit relation
\begin{equation}
    C_L = - \frac{\partial K}{\partial x } \, . \label{eq:x-of-C}
\end{equation}
 In this way, the cumulants $\mathfrak{M}_n$ not only characterize the longitudinal momentum transfer probability distribution, they explicitly determine it. Nonzero cumulants for $n \geq 3$ imply a non-Gaussian momentum transfer probability, because the resulting $\tilde{S}$ is manifestly not a quadratic function of $C_L$.

It is thus worth examining what structural properties are present in $K$ beyond its symmetries. One such property that our calculation gives us access to, which, to our knowledge, has not received much attention in the literature, is the large $n$ behavior of the longitudinal cumulants, which determines when the series~\eqref{eq:Kxv} converges. In fact, at $v=0$ the moments can be computed in closed form (see End Matter). The even cumulants scale as
\begin{equation}
    \mathfrak{M}_n(0) \sim \, (2T)^n \, \frac{n!}{n^2}, \qquad n\to\infty, \quad n \ {\rm even},
\end{equation}
giving a radius of convergence $R(v=0) = 1/(2T)$ for the series that determines $K$ in Eq.~\eqref{eq:Kxv}.  Similarly, in the ultra-relativistic limit $\gamma \gg 1$, the radius of convergence becomes $R(v) \sim 1/(4\gamma^2 T)$~\cite{DuPlessis:2026qvj}. 
 $R(v)$ contains dynamical information about the microscopic nature of the medium: Applying the same characterization to the known strongly coupled $\mathcal{N}=4$ SYM kernel~\cite{Rajagopal:2025ukd}, we obtain $R(v) \sim 1/(r_0 \gamma^\frac{3}{2} T)$ --- featuring a different exponent for $\gamma$, with $r_0 \approx 1.7972$ --- sketched in the End Matter. Note that, both in the weak and strong coupling expansions, the leading contribution to 
$R(v)$ is independent of the coupling --- coupling dependence only appears at subleading order in either expansion. 

Furthermore, in both weakly coupled gauge plasmas and this strongly coupled theory, all of the cumulants are positive, and therefore $K$ exhibits singularities at the edges of its domain of analyticity outside the interval $-v/T < x < 0$. Through Eq.~\eqref{eq:x-of-C}, these edges map to $C_L \to \pm\infty$ due to the rapid growth of $\mathfrak{M}_n$ with $n$ in all of these theories, and because the inverse Legendre transform gives $x = \partial\tilde{S}/\partial C_L$, the momentum transfer probability has exponential tails at large $|C_L|$, with its steepness in either direction set by the radius of convergence $R(v)$ of the kernel $K$ and the asymmetry required by the universal equilibration condition~\cite{Rajagopal:2025rxr}. Explicitly,
\begin{align}
    &P(k_L) \propto e^{- R(v) k_L } & & {\rm as} \,\, k_L \to \infty \, , \\
    &P(k_L) \propto e^{ +[R(v) + v/T] k_L } & &{\rm as} \,\, k_L \to -\infty \, .
\end{align}
This physical feature of the momentum transfer probability seems to be robust across all couplings (manifest in Fig.~\ref{fig:legendre} below), directly connected to $K$ having a finite, nonzero radius of convergence. Thus, the radius of convergence $R(v)$ of the Taylor expansion of $K(x,v)$ about $x=0$ is also exactly the exponential decay rate of the positive large $k_L$ tail. We therefore refer to this quantity as the \textit{tail exponent}. The asymptotic negative-$k_L$ decay rate is $R(v)+\frac{v}{T}$, fixed by the generalized equilibration condition.

As one might expect, the main driver of the exponential tails is the hard physics of the medium (in the HTL sense, i.e., $|{\bf p}|\sim T$; \textit{not} in the HQET $|{\bf k}| \sim M$ sense), manifest in the fact that the high order cumulants are set by the perturbative contribution, and not the soft physics ($|{\bf p}|\sim gT$) described by HTL resummation. Looking at the structure of the calculation, it should remain true at any order in perturbation theory that the large $n$ scaling of the cumulants, and therefore the radius of convergence, are solely influenced by the hard physics contributions. Furthermore, in an asymptotically free gauge theory, if the coupling runs with the momentum transfer, the tails of the momentum transfer should asymptotically be given by the leading order hard piece we have computed here.
\begin{figure}
    \centering
    \includegraphics[width=0.98\linewidth]{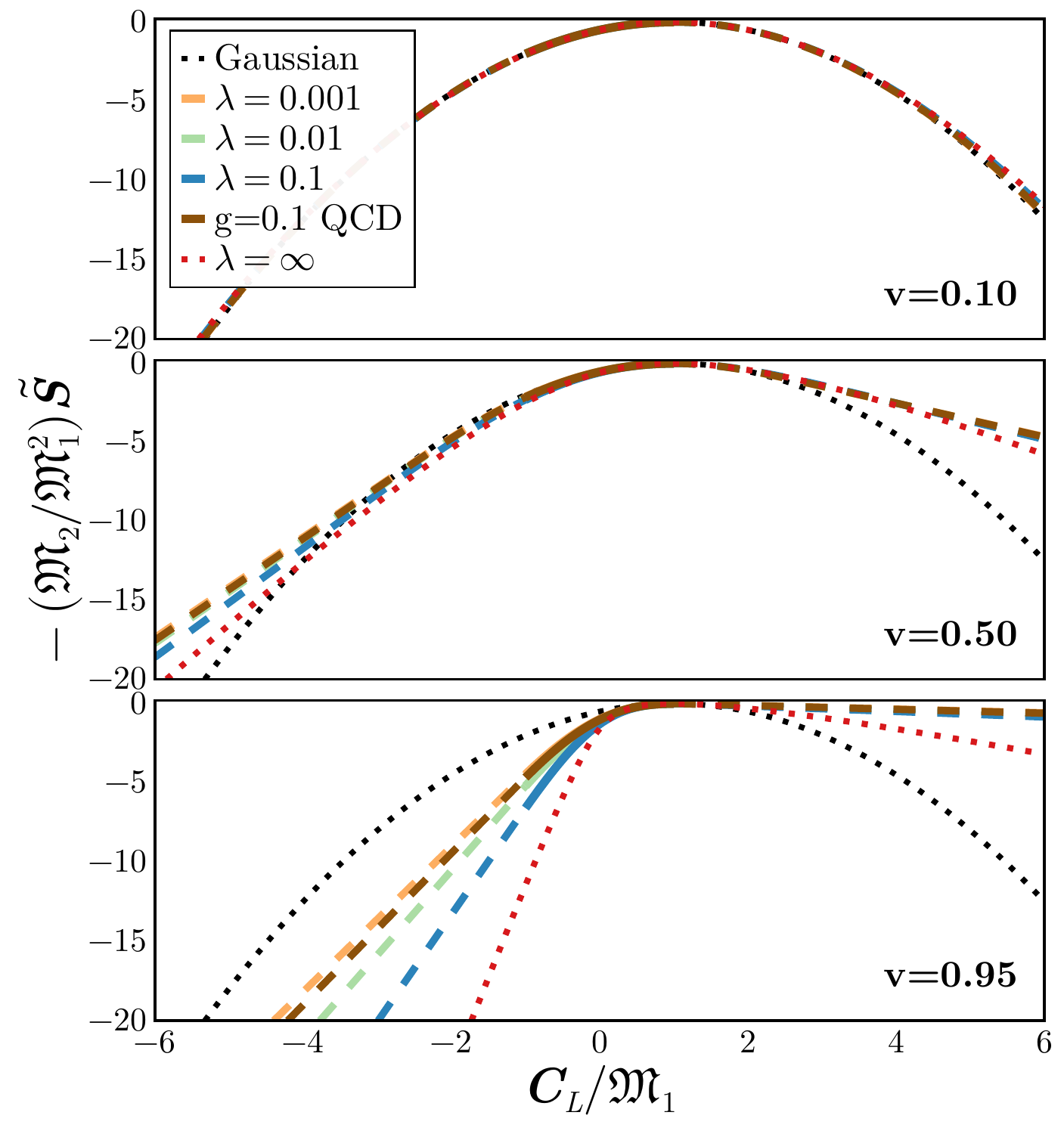}
    \caption{\label{fig:legendre}
    Normalized log-probability density for longitudinal momentum transfer, shown as $(\mathfrak{M}_2/\mathfrak{M}_1^2)\tilde S(C_L;v)$ versus $C_L/\mathfrak{M}_1$ for $v=0.10$, $0.50$, and $0.95$. Rescaling by the first two longitudinal cumulants removes the overall drag and width, so that any purely Gaussian process collapses onto the universal parabola shown by the black dotted curve; deviations from that parabola therefore isolate genuine non-Gaussian shape. Colored curves show the weak-coupling fixed-order result for $\mathcal N=4$ SYM with $\lambda=g^2N_c=10^{-3},10^{-2},10^{-1}$ together with the weak-coupling massless $N_f=3$ QCD result. Solid segments are obtained directly from the integral representation of Eq.~\eqref{eq:SFO-main} inside the interval $-v/T < x < 0$, while dashed segments are obtained by analytic continuation using the moments in Eq.~\eqref{eq:Kxv}. The red dotted curve is the strong-coupling $\mathcal N=4$ SYM result of Ref.~\cite{Rajagopal:2025ukd}. The figure shows that, once the Gaussian scale set by $\mathfrak{M}_1$ and $\mathfrak{M}_2$ is removed, weakly coupled $\mathcal N=4$ SYM, strongly coupled $\mathcal N=4$ SYM, and weakly coupled QCD all exhibit the same qualitative non-Gaussian structure.}
\end{figure}

To compare the shape of the longitudinal momentum transfer~\eqref{eq:P-H} distribution across velocities, couplings and theories, we normalize the logarithm of said distribution using its first two longitudinal cumulants and plot $(\mathfrak{M}_2/\mathfrak{M}_1^2)\tilde S(C_L;v)$ against $C_L/\mathfrak{M}_1$ in Fig.~\ref{fig:legendre}. With this normalization, the overall drag and width are scaled out, and any purely Gaussian process collapses onto a universal parabola, shown by the black dotted curve. Deviations from that benchmark therefore isolate genuine non-Gaussian structure.

At low velocity, $v=0.10$, the weak-coupling result remains close to the Gaussian benchmark in the range plotted. By $v=0.50$, however, the distribution is already visibly skewed and non-Gaussian, and by $v=0.95$ the deviation is pronounced on both sides of the extremum. Over the range $\lambda=10^{-3}$--$10^{-1}$ the normalized weakly coupled $\mathcal N=4$ curves\footnote{ We note that our calculation is based on the QCD Wilson line ${\rm P} \exp \left( i g \int A_\mu dx^\mu \right)$. A different calculation in $\mathcal{N}=4$ SYM would involve Wilson lines that also include a direct coupling of the HQ to adjoint scalars, like earlier studies of HQ diffusion in this theory have done~\cite{Chesler:2006gr,Caron-Huot:2008dyw}. However, as discussed in~\cite{Rajagopal:2025ukd}, the distinction between the two types of Wilson lines is inconsequential for the calculation of the Wilson loop~\eqref{eq:W-S} at strong coupling~\cite{Alday:2007he,Polchinski:2011im,Casalderrey-Solana:2011dxg}, which is what we aim to compare against here.} have little spread, and the weak-coupling massless $N_f=3$ QCD curve has the same qualitative shape. The strong-coupling $\mathcal N=4$ SYM result of Ref.~\cite{Rajagopal:2025ukd}, shown as the red dotted curve, exhibits the same qualitative asymmetric non-Gaussian profile. Contrary to what Fig.~\ref{fig:legendre} might suggest, we remark that as $t T \to \infty$ the bulk of the distribution is asymptotically approximated by a Gaussian distribution for any fixed $v$, as required by the central limit theorem. The tails --- the feature that this figure is designed to highlight --- characterize the rare events away from the (Gaussian) core of the distribution that become suppressed in the late time limit, but still retain a nonzero probability of occurrence. That being said, regardless of their small probability, the asymmetry of the exponential tails is in fact essential for heavy quark equilibration~\cite{Rajagopal:2025rxr}, which is evident in the fact that the Gaussian core \textit{does not} satisfy the Einstein relation. Furthermore, for a realistic heavy quark with large but finite mass traversing the medium for a finite time, the asymptotic central limit theorem regime can never be fully reached: individual large kicks -- precisely those described by the exponential tails -- can appreciably change the quark's velocity $v\to v'$, forcing subsequent samplings to be from $P(k_L;v')$ which is not well-described by the original $P(k_L;v)$. Therefore, the tails carry non-negligible dynamical importance beyond their role in ensuring equilibration.

The weak/strong agreement in $\mathcal N=4$ SYM we see in Fig.~\ref{fig:legendre}, together with the same qualitative weak-coupling structure in QCD, suggests that Gaussian core and exponential tail momentum transfer is not peculiar to weak or strong coupling, conformality, or supersymmetry, but rather a robust feature that realistic QGP should also exhibit, with its \textit{independent} Gaussian characteristics $\eta_D(v)$, $\kappa_L(v)$, and the tail exponent $R(v)$ as the primary properties of the dynamics. 

\textbf{Outlook.}
We have shown that the longitudinal momentum transfer distribution of a relativistic heavy quark in generic weakly coupled gauge theory plasmas has a Gaussian core and exponential tails, and that, in particular, weakly coupled QCD exhibits this behavior. This is the same qualitative structure previously found in strongly coupled $\mathcal N=4$ SYM theory~\cite{Rajagopal:2025ukd}. This indicates that these features might be generic across many gauge theory plasmas, regardless of their coupling strength or their matter content. In particular, neither conformality nor supersymmetry seem to play a significant role.

Isolating the strict fixed-order kernel was essential to making this structure calculable at weak coupling.
Dissecting the two types of contributions to our result --- distinguished by whether there is a logarithmic $\ln(1/g)$ enhancement --- we find that only the Gaussian characteristics of the momentum transfer probability contain logarithmically enhanced terms. Accordingly, the log-enhanced terms satisfy the Einstein relation, as had long been known~\cite{Moore:2004tg}. At finite velocity, however, the remaining fixed-order contributions control the shape of the distribution and invalidate a purely Gaussian truncation, as the universal equilibration condition~\cite{Rajagopal:2025rxr} is only satisfied through a detailed balance of all of the cumulants.
A Gaussian description is therefore not consistent with first-principles calculations for generic relativistic heavy-quark transport.

These results have direct implications for heavy-flavor phenomenology. Since non-Gaussian momentum transfer can modify heavy-flavor observables and the transport coefficients extracted from them~\cite{Das:2013kea,Scardina:2017ipo,Beraudo:2025nvq}, phenomenological descriptions based purely on Gaussian or Langevin truncations are not microscopically justified by default. Our calculation therefore motivates revisiting heavy-quark transport in the QGP with non-Gaussian kernels if one aims to extract transport coefficients that can be meaningfully compared to lattice determinations. A more realistic, \emph{non-Gaussian} description may be organized in terms of $\eta_D(v)$, $\kappa_L(v)$, and a single additional parameter $R(v)$, the tail exponent, characterizing the distribution of (rare) large momentum kicks, which sets the slopes of the asymmetric exponential tails. Whether such a description can consistently account for heavy-flavor observables as the Pb--Pb program becomes increasingly precise in Run 3 and, in the longer term, with the proposed ALICE 3 detector upgrade~\cite{ALICE:2022wwr}, while also accommodating the complementary system-size constraints that should emerge from the recently collected O--O and Ne--Ne data, should provide a sharp test.

\begin{acknowledgments}
We gratefully acknowledge helpful conversations with Dani Pablos, Krishna Rajagopal, Rachel Steinhorst, Eamonn Weitz, Urs Wiedemann, and Xiaojun Yao.
Research supported in part by the U.S.~Department of Energy, Office of Science, Office of Nuclear Physics under grant Contract Number DE-SC0011090, by grant NSF PHY-2309135 to the Kavli Institute for Theoretical Physics (KITP), and by grant 994312 from the Simons Foundation. This research has benefited from the Mitchel Postdoctoral Scholar Career Development Fund at KITP.
\end{acknowledgments}

\bibliography{main.bib}

\appendix

\section{End Matter}
\renewcommand{\theequation}{A.\arabic{equation}}
\setcounter{equation}{0}

\paragraph{Integrands.}\label{app:ints}

\begin{figure}
    \centering
    \includegraphics[]{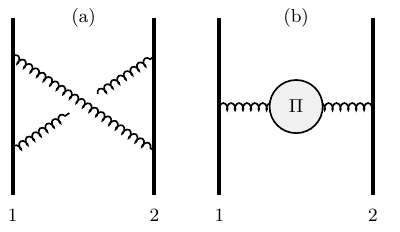}
    \caption{Contributing order-$g^4$ perturbative topologies in Feynman gauge. Panel (a) denotes the crossed one-gluon exchanges between opposite contour branches; panel (b) denotes the self-energy correction to a single exchanged gluon, including gauge, fermion, and scalar contributions.}
    \label{fig:diagrams}
\end{figure}

From perturbative finite temperature field theory one obtains
\begin{align}
    &\mathcal{I}_{\rm Hard}=\frac{N_c}{2} \int_{k_0}  \frac{e^{{\bf p} \cdot {\bf v}/T} - 1 }{|e^{k_0/T}\!-\!1| |e^{({\bf p} \cdot {\bf v}-k_0)/T} \!-\! 1| } 
\frac{\Theta \! \left( 1\! -\! u^2 \right)}{ 4 |{\bf p}|}    \nonumber \\
& \times \Bigg(  \frac{ (1- {\bf v}^2 )^2  \left| {\rm sgn}(k_0) - u \frac{{\bf v} \cdot {\bf p}}{|{\bf p}|}  \right| }{  k_0^2 \! \left[ \left(\!{\rm sgn}(k_0)\! - \!u \frac{{\bf v} \cdot {\bf p}}{|{\bf p}|} \right)^2\! - (1\!-\!u^2) \left( {\bf v}^2 \!-\! \frac{({\bf v} \cdot {\bf p})^2}{{\bf p}^2} \right)   \right]^{3/2} }\nonumber \\ & -\! \frac{4(1\!-\!{\bf v}^2) }{  {\bf p}^2 \!-\!  ( {\bf p}\! \cdot \!{\bf v} )^2   }  + \frac{8 + 4N_X }{ [{\bf p}^2\! -\!  ( {\bf p} \cdot {\bf v} )^2 ]^2 } \Bigg[ \left( k_0\! -\! |k_0| \frac{{\bf p}\! \cdot \!{\bf v}}{|{\bf p}|} u \right)^2 \nonumber \\ &\qquad\qquad\qquad\quad + \frac{k_0^2}{2} \left( {\bf v}^2\! -\! \frac{ ( {\bf p}\!\cdot\! {\bf v} )^2 }{{\bf p}^2} \right) \left( 1\! - \!u^2 \right) \Bigg] \Bigg) \nonumber \\
&  -\! 2 N_f T_F  \int_{k_0}\! \frac{ e^{k_0/T} (1\! -\! e^{- {\bf p} \cdot {\bf v}/T }) }{(e^{k_0/T}\!+\!1) (e^{(k_0-{\bf p} \cdot {\bf v})/T}\! +\! 1) }
\frac{ \Theta \! \left( 1\! -\! u^2 \right)}{ 4 |{\bf p}| } \nonumber \\ &\times \!\Bigg( \frac{1-{\bf v}^2}{ {\bf p}^2\! -\! ({\bf p}\!\cdot\! {\bf v})^2 }\!  - \!\frac{4}{[{\bf p}^2\! -\! ({\bf p}\!\cdot\! {\bf v})^2 ]^2} \Bigg[ \left( k_0\! - \!  |k_0| \frac{{\bf p}\! \cdot\! {\bf v}}{|{\bf p}|} u \right)^2 \nonumber \\ &\qquad\qquad\qquad\quad+ \frac{k_0^2}2  \left( {\bf v}^2\! -\! \frac{ ( {\bf p}\! \cdot \!{\bf v} )^2 }{{\bf p}^2} \right) \left( 1\! -\! u^2 \right) \Bigg] \Bigg),
\end{align}
where we have introduced $u$ as
\begin{equation}
	u = \frac{ {\bf p}^2 + 2 ({\bf p} \cdot {\bf v}) k_0 - ({\bf p} \cdot {\bf v})^2 }{2|{\bf p}| |k_0|} \, ,
\end{equation}
describing the cosine of the angle between a loop momentum and ${\bf p}$. The Heaviside step function implies that $ |{\bf p}| - |{\bf p} \cdot {\bf v}| < 2 |k_0| < |{\bf p}| + |{\bf p} \cdot {\bf v}| $, which in turn implies that $\mathcal{I}_{\rm Hard}$ is exponentially suppressed at large ${\bf p}$.
Note that, at this order in Feynman gauge perturbation theory, only the two classes of perturbative diagrams shown in Fig.~\ref{fig:diagrams} contribute. All other topologies are either ${\bf L}$-independent, sub-extensive in the late-time limit, or vanish identically. In other gauges, a 3-gluon vertex diagram is also present. The result is obtained by a tedious but straightforward evaluation of integrals over delta functions; further details will be presented in Ref.~\cite{DuPlessis:2026qvj}.

Using the standard HTL-resummed gluon self-energy in the single exchange propagator~\cite{Braaten:1989mz,Caron-Huot:2007cma,Laine:2016hma} gives
\begin{align}
    \mathcal{I}_{\rm Soft}=&\frac{ 2 \Gamma_E( ({\bf p} \cdot {\bf v})/|{\bf p}|  )   }{ \Sigma_E^2( p_0 = {\bf p} \cdot {\bf v} , p ) + \Gamma_E^2( ({\bf p} \cdot {\bf v})/|{\bf p}|  ) }  \nonumber \\
   +& \frac{2 {\bf v}^2 \left( 1 - \frac{({\bf p} \cdot {\bf v})^2}{{\bf p}^2 {\bf v}^2} \right) \, \Gamma_T( ({\bf p} \cdot {\bf v})/|{\bf p}| )  }{ \Sigma_T^2( p_0 = {\bf p} \cdot {\bf v}, p ) + \Gamma_T^2( ({\bf p} \cdot {\bf v})/|{\bf p}| ) }. 
\end{align}
Expanding to lowest order in the coupling yields the contribution that has been over-counted by adding the hard and soft contributions to each other and must therefore be subtracted
\begin{align}
    \mathcal{I}_{\rm Match}&=\frac{ 2 \Gamma_E( ({\bf p} \cdot {\bf v})/|{\bf p}|  )   }{ |{\bf p}|^4 }\nonumber\\&+\frac{2 {\bf v}^2 \left( 1 - \frac{({\bf p} \cdot {\bf v})^2}{{\bf p}^2 {\bf v}^2 } \right) \, \Gamma_T( ({\bf p} \cdot {\bf v})/|{\bf p}|  )  }{ |{\bf p}|^4 \left(1 -  \frac{({\bf p} \cdot {\bf v})^2}{{\bf p}^2  }  \right)^2 }, \label{eq:Imatch-def}
\end{align}
where
\begin{align}
    \Sigma_T(k_0 = k \eta, k) &= k^2 (1 - \eta^2 )\nonumber  \\ &\!\!\!\!\!\!\!\!\!\!\!+ \frac{m_D^2}{2} \left[ \eta^2+ \frac{\eta (1 - \eta^2 ) }{2} \ln \left| \frac{1 + \eta}{1 -\eta} \right| \right] \, , \\
    \Sigma_E(k_0 = k \eta, k) &= k^2 + m_D^2 \left[ 1 - \frac{ \eta }{2} \ln \left| \frac{1 + \eta}{1 - \eta} \right| \right] \, , \\
    \Gamma_T(\eta) &= \frac{\pi m_D^2}{4} \eta (1 - \eta^2) \, , \\
    \Gamma_E(\eta) &= \frac{\pi m_D^2}{2} \eta \, .
\end{align}

\paragraph{Invariance under change in separation scale.}

Choosing an arbitrary matching scale $\Lambda\sim g^0 T$ leads to Eq.~\eqref{eq:SFO-main} changing in two simple ways:
\begin{enumerate}
    \item The explicit logarithm in the first line $\ln (m_D/T)$ becomes $\ln(m_D/\Lambda)$, and
    \item the Heaviside step function in the last line $\theta(T-|{\bf p}|)$ becomes $\theta(\Lambda-|{\bf p}|)$.
\end{enumerate}
Then, differentiating with respect to $\Lambda$ gives a $1/\Lambda$ for the first and a $\delta(\Lambda-|{\bf p}|)$ for the second. Using the explicit form of $\mathcal{I}_{\rm Match}$~\eqref{eq:Imatch-def}, it is then straightforward to check that the two contributions exactly cancel. Physically, this is because the log-enhanced contributions coming from the HTL-resummed propagator can be equivalently obtained by integrating the unresummed HTL propagator from $|{\bf p}|\sim m_D$ to $|{\bf p}|\sim g^0T$ as discussed in~\cite{Moore:2004tg}.

\paragraph{Closed form of cumulants at $v=0$.}
From Eqs.~\eqref{eq:SFO-main} and~\eqref{eq:Kxv} it is clear that $\mathfrak{M}_n$ for $n\ge4$ is fully determined by the hard contribution. 
Introducing $k\equiv 2|k_0|/p\ge1$, exploiting the symmetries of the integrands, it is straightforward to show that
\begin{align}
    &K_{\rm hard}(x,0)
    = \\
    &-\frac{g^4 C_F}{64\pi^3 T}
    \int_0^\infty \!\!\! dp \! \int_{-1}^1 \!\! d\mu \, \sinh^2\!\left(\frac{p\mu x}{2}\right) \! \int_1^\infty \!\!\! dk \nonumber \\ & \,\,\, \Bigg\{
    \frac{N_c [2k^{-2}-2+ (1 + N_X/2) k^2 ] }{\sinh^2\!\left(\frac{pk}{4T}\right)}
    + \frac{2 T_F N_f (k^2-1) }{\cosh^2\!\left(\frac{pk}{4T}\right)}
    \Bigg\}. \nonumber
    \label{eq:Khard-EM}
\end{align}
Because the cumulants are given by derivatives of $K$ as $\mathfrak{M}_n = - \partial^n K/\partial x^n|_{x=0}$, and the only scale in the problem is $T$, all that is needed is to evaluate the integrals
\begin{align}
    \tilde{\mathcal J}^{(\sinh)}_{n,m}
    &\equiv
    \int_0^\infty \!\!\! dp \!
    \int_1^\infty \!\!\! dk
    \frac{ p^n k^m}{\sinh^2\!\left(\frac{pk}{4}\right)} = \frac{2^{n+3}n!\,\zeta(n)}{(n-m)} \, ,\\
    \tilde{\mathcal J}^{(\cosh)}_{n,m}
    &\equiv
    \int_0^\infty \!\!\! dp \!
    \int_1^\infty \!\!\! dk
    \frac{p^n k^m}{\cosh^2\!\left(\frac{pk}{4}\right)} = (1-2^{1-n} ) \tilde{\mathcal{J}}^{(\sinh)}_{n,m} \, .
\end{align}
Using these to evaluate the cumulants yields
\begin{widetext}
\begin{equation}
    \boxed{
    \mathfrak M_n(0)
    =
    \frac{3\,g^2 C_F m_D^2\,2^{\,n-1}T^{\,n-2}(n-1)!\,\zeta(n)}
    {\pi^3 (n+1)\!\left(N_c+\frac{N_cN_X}{2}+T_F N_f\right)}
    \left[
    \frac{N_c\!\left(N_X n(n+2)+2(n^2-2n+8)\right)}{8(n-2)(n+2)}
    +\frac{(1-2^{1-n})\,T_F N_f}{n-2}
    \right]
    }
    \label{eq:Mn-v0-closed-EM}
\end{equation}    
\end{widetext}
for all even $n\ge4$, while $\mathfrak M_n(0)=0$ for odd $n$. The explicit
$(n-2)^{-1}$ pole signals the special infrared sensitivity of the diffusion
coefficient, and is precisely why the $n=2$ case must be treated separately.

\paragraph{HQ diffusion coefficient at $v=0$.}
The heavy quark momentum diffusion coefficient is determined in terms of $\mathfrak{M}_2(0)$ as
\begin{equation}
    \kappa \equiv T\,\mathfrak M_2(0)\, .
\end{equation}
We set $N_X = 0$ for easier comparison with the QCD literature. To evaluate $\mathfrak{M}_2(0)$, it is convenient to analytically continue the hard result
in Eq.~\eqref{eq:Mn-v0-closed-EM} in $n$, combine it with the quadratic
HTL$-$matching part of Eq.~\eqref{eq:SFO-main} also analytically continued in $n$, and then take the limit $n\to2$. 
The poles cancel in the sum, as they must. We obtain
\begin{equation}
    \boxed{
    \kappa
    =
    \frac{g^2 C_F m_D^2 T}{6\pi}
    \Bigg[
        \ln \left(\frac{2T}{m_D}\right)
        +\frac{T_F N_f \ln 2}{N_c+T_F N_f}
         + \xi
    \Bigg]
    }
    \label{eq:kappa-v0-EM}
\end{equation}
with $\xi = \tfrac{1}{2}-\gamma_E+\tfrac{\zeta'(2)}{\zeta(2)}$, in agreement with the known leading-order heavy quark diffusion coefficient~\cite{Moore:2004tg,Burnier:2010rp}.

\paragraph{$R(v)$ in the ultra-relativistic limit at weak coupling.}
In the $v\to1$ limit the dominant contribution to the large-$n$ cumulants comes from nearly collinear corners of the hard integrand, where ${\bf p} \cdot {\bf v} \simeq |{\bf p}|$, asymptotically pinched against the kinematic boundary imposed by $\Theta(1-u^2)$, which sets the boundaries of the $k_0$ integral to be $k_0 > (|{\bf p}| + {\bf p} \cdot {\bf v}) /2$ and $k_0 < -(|{\bf p}| - {\bf p} \cdot {\bf v}) /2$. To isolate these corners it is convenient to introduce
\begin{equation}
    \ell \equiv \frac{k_0}{p},
    \qquad
    q \equiv (1-v^2)\frac{p}{T}.
\end{equation}
The variable $\ell$ makes the kinematic support asymptotically independent of $p$, since the hard-region bounds are linear in $p$, while $q$ removes the leading velocity dependence of the singularity driven by the difference between ${\bf p} \cdot {\bf v}$ and $|{\bf p}|$. Since $p=\gamma^2 T\,q$, the $n$th moment weight contributes a factor $(\gamma^2 T)^n$, which is the origin of the $\gamma^{2n}T^n$ scaling quoted in the main text. Schematically, the remaining $q$-integral produces the factorial growth through $\int dq\, q^n/(e^q-1)=n!\zeta(n+1)$. A more detailed account will be given in~\cite{DuPlessis:2026qvj}. The conclusion of this analysis is that as $\gamma \to \infty$ one has
\begin{equation}
    R(v) \sim 1/(4 \gamma^2 T) \, .
\end{equation}

\paragraph{$R(v)$ in the ultra-relativistic limit at strong coupling.}

In light of the Legendre transform map $x = \partial \tilde{S}/\partial C_L$, and because the explicit expression for $\tilde{S}$ is known~\cite{Rajagopal:2025ukd}, the radius of convergence of $K(x)$ in strongly coupled $\mathcal{N}=4$ SYM can be determined by directly calculating the derivative of $\tilde{S}$ with respect to $C_L$ at large $|C_L|$ (considering both positive and negative $C_L$ separately). Going through this calculation, discussed in more detail in~\cite{DuPlessis:2026qvj,Rajagopal:2026urq}, one finds
\begin{equation}
    r_0 = \frac{8 \sqrt{2} \Gamma(7/4)^2}{3\sqrt{\pi}} \approx 1.7972 \, .
\end{equation}

\clearpage
\onecolumngrid

\setcounter{section}{0}
\setcounter{subsection}{0}
\setcounter{equation}{0}
\setcounter{figure}{0}
\setcounter{table}{0}

\renewcommand{\thesection}{S\arabic{section}}
\renewcommand{\thesubsection}{S\arabic{section}.\arabic{subsection}}
\renewcommand{\theequation}{S\arabic{equation}}
\renewcommand{\thefigure}{S\arabic{figure}}
\renewcommand{\thetable}{S\arabic{table}}

\begin{center}
    {\Large\bfseries Supplemental Material}\\[0.5em]
    {\normalsize for}\\[0.5em]
    {\large\itshape Heavy Quark Transport is Non-Gaussian Beyond Leading Log}
\end{center}

\vspace{1em}

\section{Fixed order soft contribution}\label{app:fixedorder}

In this appendix we derive the fixed-order soft contribution quoted in Eq.~\eqref{eq:SFO-main}. We begin from
\begin{equation}
    - \frac{g^2 C_F}{T} \int_{\bf p } \frac{e^{i {\bf p} \cdot {\bf L} } - 1}{1 - e^{- {\bf p} \cdot {\bf v}/T } }\theta(T-|{\bf p}|)\mathcal{I}_{\rm Soft}=
    - \frac{g^2 C_F }{2}\int_{{\bf p}<T} {\bf p} \cdot i{\bf L} \,\,{\bf p} \cdot\left(i{\bf L} +\frac{{\bf v}}{T} \right)\frac{\mathcal{I}_{\rm Soft}}{{\bf p} \cdot{\bf v}}
    +\mathcal{O}(g^6),\label{eq:soft-pre-radial-app}
\end{equation}
which follows from the oddness of $\mathcal{I}_{\rm Soft}$ under ${\bf p}\to-{\bf p}$, and higher orders will cancel against the matched integrand at the relevant order.

We choose the polar axis along ${\bf v}$ and define
\begin{equation*}
    \hat{\bf v}\equiv \frac{{\bf v}}{v},\quad
    L_\parallel\equiv {\bf L}\cdot\hat{\bf v},\quad
    L_\perp^2\equiv {\bf L}^2-L_\parallel^2,\quad
    \mu\equiv \frac{{\bf p}\cdot{\bf v}}{|{\bf p}|\,v},\quad
    \eta\equiv \mu v.
\end{equation*}
After averaging over the azimuthal angle,
\begin{equation}
\int_0^{2\pi}\frac{d\phi}{2\pi}\,
\frac{{\bf p}\cdot i{\bf L}\;\;{\bf p}\cdot\left(i{\bf L}+\frac{{\bf v}}{T}\right)}
{{\bf p}\cdot{\bf v}}=
\frac{|{\bf p}|}{v}
\left[
\mu^2\, iL_\parallel\left(iL_\parallel+\frac{v}{T}\right)
+\frac{1-\mu^2}{2}(iL_\perp)^2
\right].
\label{eq:azavg-app}
\end{equation}
Using the evenness of the reduced integrand, Eq.~\eqref{eq:soft-pre-radial-app} becomes
\begin{align}
    &- \frac{g^2 C_F}{T} \int_{\bf p } \frac{e^{i {\bf p} \cdot {\bf L} } - 1}{1 - e^{- {\bf p} \cdot {\bf v}/T } }\theta(T-|{\bf p}|)\mathcal{I}_{\rm Soft}=\nonumber\\
    &
    -\frac{g^2 C_F }{(2\pi)^2}
    \int_0^1 d\mu
    \int_0^T dp\, p^3
    \Bigg[
    \mu^2 iL_\parallel\left(iL_\parallel+\frac{v}{T}\right)
    +\frac{1-\mu^2}{2}(iL_\perp)^2
    \Bigg]\Bigg[
    \frac{2\Gamma_E(\eta)}{\eta\left(\Sigma_E^2+\Gamma_E^2\right)}
    +\frac{2v^2(1-\mu^2)\Gamma_T(\eta)}{\eta\left(\Sigma_T^2+\Gamma_T^2\right)}
    \Bigg]
    +\mathcal{O}(g^6),
    \label{eq:soft-radial-app}
\end{align}
where $\Sigma_{E,T}$ and $\Gamma_{E,T}$ are evaluated at $p_0={\bf p}\cdot{\bf v}=p\,\eta$.

For the electric sector we write
\begin{equation}
    \Sigma_E = p^2 + m_D^2 a_E(\eta),\qquad
    \Gamma_E = m_D^2 b_E(\eta),
\end{equation}
with
\begin{equation}
    a_E(\eta)\equiv 1-\frac{\eta}{2}\ln\left(\frac{1+\eta}{1-\eta}\right),\qquad
    b_E(\eta)\equiv \frac{\pi}{2}\eta.
\end{equation}
For the transverse sector,
\begin{equation*}
    \Sigma_T=(1-\eta^2)\left[p^2+m_D^2 a_T(\eta)\right],\,
    \Gamma_T=(1-\eta^2)m_D^2 b_T(\eta),
\end{equation*}
with
\begin{equation}
    a_T(\eta)\equiv \frac{1}{2}\left[\frac{\eta^2}{1-\eta^2}+\frac{\eta}{2}\ln\left(\frac{1+\eta}{1-\eta}\right)\right],\,
    b_T(\eta)\equiv \frac{\pi}{4}\eta.
\end{equation}
It is useful to define
\begin{align}
    I(a,b;m_D)
    &\equiv
    \int_0^T dp\,\frac{p^3}{(p^2+a\,m_D^2)^2+(b\,m_D^2)^2}\label{eq:Iab-app}\\
    &=
    \frac14\ln\left(
    \frac{T^4+2a m_D^2 T^2+(a^2+b^2)m_D^4}{(a^2+b^2)m_D^4}
    \right)
    -\frac{a}{2b}\left[
    \arctan\left(\frac{T^2+a m_D^2}{b m_D^2}\right)
    -\arctan\left(\frac{a}{b}\right)\right].
\end{align}
We note that it is essential to use the correct branch of the $\arctan$.

Since $2\Gamma_E/\eta=\pi m_D^2$, one finds
\begin{equation}
    \int_0^T dp\, p^3\frac{2\Gamma_E(\eta)}{\eta\left(\Sigma_E^2+\Gamma_E^2\right)}=
    \pi m_D^2\, I\left(a_E(\eta),b_E(\eta);m_D\right).
\end{equation}
Similarly,
\begin{equation}
    \int_0^T dp\, p^3\frac{2v^2(1-\mu^2)\Gamma_T(\eta)}{\eta\left(\Sigma_T^2+\Gamma_T^2\right)}=
    \frac{\pi m_D^2}{2}\,
    \frac{v^2(1-\mu^2)}{1-\eta^2}\,
    I\left(a_T(\eta),b_T(\eta);m_D\right).
\end{equation}
Substituting into Eq.~\eqref{eq:soft-radial-app}, we obtain
\begin{align}
    &- \frac{g^2 C_F}{T} \int_{\bf p } \frac{e^{i {\bf p} \cdot {\bf L} } - 1}{1 - e^{- {\bf p} \cdot {\bf v}/T } }\theta(T-|{\bf p}|)\mathcal{I}_{\rm Soft}\label{eq:soft-after-radial-app}\\
    &=
    \!-\frac{g^2 C_F  m_D^2}{4\pi}\!\!
    \int_0^1 \!\!d\mu
    \Bigg[
    \mu^2 iL_\parallel\!\left(iL_\parallel\!+\!\frac{v}{T}\right)
    \!+\!\frac{1\!-\!\mu^2}{2}(iL_\perp)^2
    \Bigg]\!
    \Bigg[
    I\!\left(a_E(\eta),b_E(\eta);m_D\!\right)
    \!+\!\frac{v^2(1\!-\!\mu^2)}{2(1\!-\!\eta^2)}
    I\!\left(a_T(\eta),b_T(\eta);m_D\right)\!
    \Bigg]
    \!+\!\mathcal{O}(g^6).\nonumber
\end{align}

The crucial point is that the expansion in $m_D/T$ must be taken only after the radial integral has been done. Expanding Eq.~\eqref{eq:Iab-app} at fixed $a$ and $b$ gives
\begin{equation}
    I(a,b;m_D)
    =
    -\ln\left(\frac{m_D}{T}\right)
    -\frac14\ln\left(a^2+b^2\right)-\frac{a}{2b}\arctan\left(\frac{b}{a}\right)
    +\mathcal{O}\left(\frac{m_D^2}{T^2}\right).
    \label{eq:Iab-smallmD-app}
\end{equation}
Because Eq.~\eqref{eq:soft-after-radial-app} already carries an overall factor of $m_D^2\sim g^2T^2$, the omitted terms contribute only at $\mathcal{O}(g^6)$. Defining
\begin{equation}
    \Phi(a,b)\equiv
    \frac14\ln\left(a^2+b^2\right)
    +\frac{a}{2b}\arctan\left(\frac{b}{a}\right),
\end{equation}
we obtain
\begin{align}
    &- \frac{g^2 C_F}{T} \int_{\bf p } \frac{e^{i {\bf p} \cdot {\bf L} } - 1}{1 - e^{- {\bf p} \cdot {\bf v}/T } }\theta(T-|{\bf p}|)\mathcal{I}_{\rm Soft}\nonumber\\
    =&\frac{g^2 C_F m_D^2}{4\pi}\ln\left(\frac{m_D}{T}\right)
    \Bigg[
    \frac{\left(v^2-1\right) \arctanh(v)+v}{2 v^3}\, iL_\parallel\left(iL_\parallel+\frac{v}{T}\right)
    +\frac{3 v^3+\left(1-v^2\right)^2 \arctanh(v)-v}{4 v^3}(iL_\perp)^2
    \Bigg]\label{eq:soft-FO-main}\\
    +&\frac{g^2 C_F  m_D^2}{4\pi}
    \int_0^1 d\mu
    \Bigg[
    \mu^2\, iL_\parallel\left(iL_\parallel+\frac{v}{T}\right)
    +\frac{1-\mu^2}{2}(iL_\perp)^2
    \Bigg]
    \Bigg[
    \Phi\left(a_E(\mu v),b_E(\mu v)\right)
    +\frac{v^2(1-\mu^2)}{2(1-\eta^2)}
    \Phi\left(a_T(\mu v),b_T(\mu v)\right)
    \Bigg]
    +\mathcal{O}(g^6).\nonumber
\end{align}

\subsubsection{Closed-form coefficients of the logarithmic term}
The coefficient multiplying $\ln(m_D/T)$ is obtained by performing the $\mu$-integrals
\begin{align}
    \int_0^1 d\mu\mu^2
    \left[
    1+\frac{v^2(1-\mu^2)}{2(1-v^2\mu^2)}
    \right]
    &=
    \frac{(v^2-1)\arctanh(v)+v}{2v^3},\\
    \int_0^1 d\mu\,\frac{1-\mu^2}{2}
    \left[
    1+\frac{v^2(1-\mu^2)}{2(1-v^2\mu^2)}
    \right]
    &=
    \frac{3v^3+(1-v^2)^2\arctanh(v)-v}{4v^3},
\end{align}
which yields Eq.~\eqref{eq:soft-FO-main}.

\section{$v\to0$ limit of the diffusion coefficient}

In this appendix we verify that the second moment of the fixed-order kernel
reproduces the known weak-coupling heavy-quark diffusion coefficient at
${\bf v}={\bf 0}$.
At zero velocity there is no preferred spatial direction, so we may set
$L_\perp=0$ and write $L\equiv L_\parallel$.
We define the one-component diffusion coefficient by
\begin{equation}
    \kappa \equiv T\,\mathfrak M_2(0)
    =
    -\,T\left.\frac{d^2}{dx^2}
    K(x,0)\right|_{x=0},
\end{equation}
with $K(x,0)=S_{\rm FO}({\bf L}=-ix\hat{\bf e};{\bf v}={\bf 0})$.

The logarithmically enhanced contribution comes directly from the first line of
Eq.~\eqref{eq:SFO-main}.
Using
\begin{equation}
    \lim_{v\to0}
    \frac{(v^2-1)\arctanh(v)+v}{2v^3}
    =\frac13,
\end{equation}
we obtain
\begin{equation}
    \kappa_{\log}
    =
    \frac{g^2 C_F m_D^2 T}{6\pi}
    \ln\left(\frac{T}{m_D}\right).
    \label{eq:kappa-log-v0}
\end{equation}

We now turn to the remaining $\mathcal O(g^4)$ contribution.
For compactness in this appendix we write
\begin{equation}
    D \equiv N_c+\frac{N_cN_X}{2}+T_F N_f.
\end{equation}
Taking the $v\to0$ limit of the hard part of the last line of
Eq.~\eqref{eq:SFO-main}, introducing
\begin{equation}
    k\equiv \frac{2|k_0|}{p}\ge 1,
\end{equation}
and differentiating twice with respect to $x$, one finds
\begin{align}
    \kappa_{\rm hard}
    =
    \frac{g^2 C_F m_D^2}{4\pi T^2 D}
    \int_0^\infty dp \int_1^\infty dk \Bigg\{&
    N_c\frac{p^2}{32\pi^2}\,
   \,
    \frac{4/k^2-4+(2+N_X)k^2}{\sinh^2\left(\frac{pk}{4T}\right)}
    +T_F N_f\frac{p^2}{8\pi^2}
    \frac{k^2-1}{\cosh^2\left(\frac{pk}{4T}\right)}
    \Bigg\}.
    \label{eq:kappa-hard-v0}
\end{align}
Here the integral over $\mu$ has already been carried out, using
$\int_{-1}^{1}d\mu\,\mu^2=2/3$.

It is convenient to treat the first bosonic term separately.
Performing the $p$ integral first,
\begin{equation}
    \int_0^\infty dp\,
    \frac{p^2}{\sinh^2\left(\frac{pk}{4T}\right)}
    =
    \frac{32\pi^2 T^3}{3k^3},
\end{equation}
so that
\begin{equation}
    \int_0^\infty dp\int_1^\infty dk\,
    \frac{p^2}{32\pi^2}\,
    \frac{N_c}{D}\,
    \frac{4/k^2-4}{\sinh^2\left(\frac{pk}{4T}\right)}=
    \frac{N_c T^3}{3D}
    \int_1^\infty dk
    \left(\frac{4}{k^5}-\frac{4}{k^3}\right)
    =
    -\,\frac{N_c}{3D}\,T^3.
    \label{eq:kappa-hard-first-piece}
\end{equation}

For the remaining two terms we do the $k$ integrals exactly.
The bosonic integral is
\begin{align}
    \int_1^\infty dk\,
    \frac{k^2}{\sinh^2\left(\frac{pk}{4T}\right)}
    =
    &\frac{4T}{p}
    \left(\coth\frac{p}{4T}-1\right)
    -\frac{32T^2}{p^2}
    \ln\left(1-e^{-\frac{p}{2T}}\right)+\frac{64T^3}{p^3}
    \Li\left(2,e^{-\frac{p}{2T}}\right),
    \label{eq:kappa-bosonic-k-integral}
\end{align}
while for the fermionic contribution it is important to keep the
combination $(k^2-1)$ together:
\begin{align}
    \int_1^\infty dk\,
    \frac{k^2-1}{\cosh^2\left(\frac{pk}{4T}\right)}
    =
    \frac{32T^2}{p^2}
    \ln\left(1+e^{-\frac{p}{2T}}\right)
    -\frac{64T^3}{p^3}
    \Li\left(2,-e^{-\frac{p}{2T}}\right).
    \label{eq:kappa-fermionic-k-integral}
\end{align}
The absence of a term proportional to
$(1-\tanh(p/4T))$ in Eq.~\eqref{eq:kappa-fermionic-k-integral}
is the crucial simplification: it is precisely the cancellation that is lost if
one incorrectly splits the fermionic integrand into separate $k^2$ and $-1$
pieces too early.

Combining Eqs.~\eqref{eq:kappa-hard-first-piece},
\eqref{eq:kappa-bosonic-k-integral}, and
\eqref{eq:kappa-fermionic-k-integral}, we obtain
\begin{align}
    \kappa_{\rm hard}
    =
    \frac{g^2 C_F m_D^2}{4\pi T^2}
    \Bigg[
    -\frac{N_c}{3D}\,T^3
    +\int_0^\infty dp \Bigg\{&
    \frac{N_c+\frac{N_cN_X}{2}}{D}
    \Bigg[
    \frac{pT}{4\pi^2}
    \left(\coth\frac{p}{4T}-1\right)
    -\frac{2T^2}{\pi^2}
    \ln\left(1-e^{-\frac{p}{2T}}\right)
    +\frac{4T^3}{\pi^2 p}
    \Li\left(2,e^{-\frac{p}{2T}}\right)
    \Bigg]
    \nonumber\\
    &\quad
    +\frac{T_F N_f}{D}
    \Bigg[
    \frac{4T^2}{\pi^2}
    \ln\left(1+e^{-\frac{p}{2T}}\right)
    -\frac{8T^3}{\pi^2 p}
    \Li\left(2,-e^{-\frac{p}{2T}}\right)
    \Bigg]
    \Bigg\}
    \Bigg].
    \label{eq:kappa-hard-after-k}
\end{align}

The remaining quadratic terms in Eq.~\eqref{eq:SFO-main} come from the second
line and the explicit matching subtraction in the third line.
These must be combined with the infrared-sensitive part of
Eq.~\eqref{eq:kappa-hard-after-k} before the final $p$ integral is carried out.
The correct combination is
\begin{align}
    \kappa-\kappa_{\log}
    =
    \frac{g^2 C_F m_D^2}{4\pi T^2}
    \Bigg[
    I_{\rm elem}
    +T^3\int_0^\infty \frac{dp}{p}\Bigg\{&
    \frac{4}{\pi^2}
    \frac{N_c+\frac{N_cN_X}{2}}{D}
    \Li\left(2,e^{-\frac{p}{2T}}\right) -\frac{8}{\pi^2}
    \frac{T_F N_f}{D}
    \Li\left(2,-e^{-\frac{p}{2T}}\right)
    \nonumber\\
    &\;
    +\frac{2}{3}\,
    \theta(T-p)
    \left(
        \frac{p^4}{(p^2+m_D^2)^2}-1
    \right)
    \Bigg\}
    \Bigg],
    \label{eq:kappa-combined-v0}
\end{align}
where the elementary contribution is now
\begin{align}
    I_{\rm elem}
    =
    -\frac{N_c}{3D}\,T^3&+\frac{N_c+\frac{N_cN_X}{2}}{D}
    \Bigg[
    \int_0^\infty dp\,
    \frac{pT}{4\pi^2}
    \left(\coth\frac{p}{4T}-1\right)
    -\int_0^\infty dp\,
    \frac{2T^2}{\pi^2}
    \ln\left(1-e^{-\frac{p}{2T}}\right)
    \Bigg]
    \nonumber\\
    &
    +\frac{T_F N_f}{D}
    \int_0^\infty dp\,
    \frac{4T^2}{\pi^2}
    \ln\left(1+e^{-\frac{p}{2T}}\right).
    \label{eq:Ielem-corrected}
\end{align}
These integrals are simply:
\begin{align}
    \int_0^\infty dp\,
    \frac{pT}{4\pi^2}
    \left(\coth\frac{p}{4T}-1\right)
    &=\frac{T^3}{3},
    \\
    -\int_0^\infty dp\,
    \frac{2T^2}{\pi^2}
    \ln\left(1-e^{-\frac{p}{2T}}\right)
    &=\frac{2T^3}{3},
    \\
    \int_0^\infty dp\,
    \frac{4T^2}{\pi^2}
    \ln\left(1+e^{-\frac{p}{2T}}\right)
    &=\frac{2T^3}{3},
\end{align}
and therefore
\begin{equation}
    I_{\rm elem}
    =
    T^3\left[
    \frac{2}{3}
    +\frac{N_cN_X}{6D}
    \right].
    \label{eq:Ielem-corrected-result}
\end{equation}

The remaining one-dimensional integral is most conveniently handled by a Mellin
transform.
Since extending the support of the last term in
Eq.~\eqref{eq:kappa-combined-v0} from $p<T$ to all $p$ changes the result only
at $\mathcal O(g^6)$, we may equivalently write
\begin{align}
    \kappa-\kappa_{\log}
    =
    \frac{g^2 C_F m_D^2}{4\pi T^2}
    \Bigg[
    I_{\rm elem}
    +T^3\int_0^\infty \frac{dp}{p}\Bigg\{&
    \frac{4}{\pi^2}
    \frac{N_c+\frac{N_cN_X}{2}}{D}
    \Li\left(2,e^{-\frac{p}{2T}}\right)
    -\frac{8}{\pi^2}
    \frac{T_F N_f}{D}
    \Li\left(2,-e^{-\frac{p}{2T}}\right)
    \nonumber\\
    &\;
    +\frac{2}{3}
    \left(
        \frac{p^4}{(p^2+m_D^2)^2}-1
    \right)
    \Bigg\}
    \Bigg].
    \label{eq:kappa-combined-v0-inf}
\end{align}

Introducing a regulator $(p/2T)^s$ and using
\begin{align}
    \int_0^\infty dx\,x^{s-1}\Li\left(2,e^{-x}\right)
    &=\Gamma(s)\zeta(s+2),
    \\
    \int_0^\infty dx\,x^{s-1}\Li\left(2,-e^{-x}\right)
    &=-\Gamma(s)\eta(s+2),
\end{align}
together with
\begin{equation}
    \int_0^\infty dp\,p^{s-1}
    \left(
        \frac{p^4}{(p^2+m_D^2)^2}-1
    \right)
    =
    -\frac{m_D^s}{4}\,\pi(s+2)\csc\left(\frac{\pi s}{2}\right),
\end{equation}
and
\begin{equation}
    \eta(s+2)=\bigl(1-2^{-s-1}\bigr)\zeta(s+2),
\end{equation}
we obtain, after expanding around $s=0$,
\begin{align}
    &T^3\int_0^\infty \frac{dp}{p}\Bigg\{
    \frac{4}{\pi^2}
    \frac{N_c+\frac{N_cN_X}{2}}{D}
    \Li\left(2,e^{-\frac{p}{2T}}\right)
    -\frac{8}{\pi^2}
    \frac{T_F N_f}{D}
    \Li\left(2,-e^{-\frac{p}{2T}}\right)+\frac{2}{3}
    \left(
        \frac{p^4}{(p^2+m_D^2)^2}-1
    \right)
    \Bigg\}=
    \nonumber\\
    &
    -\frac{T^3}{3}
    \Bigg[
        1+2\gamma_E
        -2\ln\left(\frac{2T}{m_D}\right)
        -2\frac{T_F N_f}{D}\ln 2
        -2\frac{\zeta'(2)}{\zeta(2)}
    \Bigg].
    \label{eq:kappa-mellin-result}
\end{align}

Collecting Eqs.~\eqref{eq:kappa-log-v0},
\eqref{eq:Ielem-corrected-result}, and
\eqref{eq:kappa-mellin-result}, we finally obtain
\begin{equation}
    \kappa
    =
    \frac{g^2 C_F m_D^2 T}{6\pi}
    \Bigg[
        \frac12
        +\frac{N_c N_X}{4\left(N_c+\frac{N_cN_X}{2}+T_F N_f\right)}
        -\gamma_E
        +\ln\left(\frac{2T}{m_D}\right)
        +\frac{T_F N_f}{N_c+\frac{N_cN_X}{2}+T_F N_f}\ln 2
        +\frac{\zeta'(2)}{\zeta(2)}
    \Bigg].
    \label{eq:kappa-v0-final}
\end{equation}

For QCD, $N_X=0$.
Our $\kappa$ is the diffusion coefficient for a single Cartesian component.
To compare with the convention of
Refs.~\cite{Moore:2004tg,Burnier:2010rp}, where the correlator is summed over
the three spatial directions, one multiplies Eq.~\eqref{eq:kappa-v0-final} by
$3$.

\section{$v\to0$ limit of higher moments}

In this appendix we derive the higher projected longitudinal moments at zero velocity.
Since there is no preferred direction for ${\bf v}={\bf 0}$, we again specialize to
$L_\perp=0$ and write $L\equiv L_\parallel$.
We also use the longitudinal kernel
\begin{equation}
    K(x,0)=S_{\rm FO}({\bf L}=-ix\hat{\bf e};{\bf v}={\bf 0})
    =-\sum_{n=1}^\infty \frac{\mathfrak M_n(0)}{n!}\,x^n.
\end{equation}
Therefore
\begin{equation}
    \mathfrak M_n(0)
    =
    -\left.\frac{d^n K(x,0)}{dx^n}\right|_{x=0}.
\end{equation}

For $n\ge4$, only the hard contribution is needed.
Indeed, the first two lines of Eq.~\eqref{eq:SFO-main} are manifestly quadratic in $L$,
and the explicit matching subtraction in the last line is also quadratic in $L$.
Hence they do not contribute to $\mathfrak M_n(0)$ for $n\ge4$.
Only the hard part of the last line contributes.

Taking the $v\to0$ limit of the hard integrand, and introducing
\begin{equation}
    k\equiv \frac{2|k_0|}{p}\ge1,
\end{equation}
exactly as in the previous appendix, one finds
\begin{align}
    K_{\rm hard}(x,0)
    =-\frac{g^2 C_F m_D^2}{4\pi T^3}
    \int_0^\infty dp\int_{-1}^1d\mu\int_1^\infty dk\Bigg\{&
    \frac{3}{32\pi^2}
    \frac{N_c}{N_c+\frac{N_cN_X}{2}+T_F N_f}\,
    \frac{\sinh^2\left(\frac{p\mu x}{2}\right)}{\sinh^2\left(\frac{pk}{4T}\right)}
    \left(\frac{4}{k^2}-4\right)
    \nonumber\\
    &\;
    -\frac{3}{8\pi^2}
    \frac{T_F N_f}{N_c+\frac{N_cN_X}{2}+T_F N_f}\,
    \frac{\sinh^2\left(\frac{p\mu x}{2}\right)}{\cosh^2\left(\frac{pk}{4T}\right)}
    \nonumber\\
    &\;
    +k^2\Bigg[
    \frac{3}{16\pi^2}
    \frac{N_c+\frac{N_cN_X}{2}}{N_c+\frac{N_cN_X}{2}+T_F N_f}\,
    \frac{\sinh^2\left(\frac{p\mu x}{2}\right)}{\sinh^2\left(\frac{pk}{4T}\right)}
    \nonumber\\
    &\hspace{1.8cm}
    +\frac{3}{8\pi^2}
    \frac{T_F N_f}{N_c+\frac{N_cN_X}{2}+T_F N_f}\,
    \frac{\sinh^2\left(\frac{p\mu x}{2}\right)}{\cosh^2\left(\frac{pk}{4T}\right)}
    \Bigg]
    \Bigg\}.
    \label{eq:Khard-v0}
\end{align}

Because $\sinh^2\left(\frac{p\mu x}{2}\right)$ is an even function of both $x$ and $\mu$,
all odd moments vanish:
\begin{equation}
    \mathfrak M_{2\ell+1}(0)=0.
\end{equation}
For even $n$,
\begin{equation}
    \left.\frac{d^n}{dx^n}\sinh^2\left(\frac{p\mu x}{2}\right)\right|_{x=0}
    =\frac12 (p\mu)^n,
\end{equation}
so for even $n\ge4$ we obtain
\begin{align}
    \mathfrak M_n(0)
    =
    \frac{g^2 C_F m_D^2}{4\pi T^3}
    \Bigg\{&
    \frac{3}{32\pi^2}
    \frac{N_c}{N_c+\frac{N_cN_X}{2}+T_F N_f}\,
    4\Big(
        \mathcal J^{(\sinh)}_{n,-2}(T)-\mathcal J^{(\sinh)}_{n,0}(T)
    \Big)
    \nonumber\\
    &\;
    +\frac{3}{16\pi^2}
    \frac{N_c+\frac{N_cN_X}{2}}{N_c+\frac{N_cN_X}{2}+T_F N_f}\,
    \mathcal J^{(\sinh)}_{n,2}(T)
    \nonumber\\
    &\;
    +\frac{3}{8\pi^2}
    \frac{T_F N_f}{N_c+\frac{N_cN_X}{2}+T_F N_f}\,
    \Big(
        \mathcal J^{(\cosh)}_{n,2}(T)-\mathcal J^{(\cosh)}_{n,0}(T)
    \Big)
    \Bigg\},
    \label{eq:Mn-v0-J}
\end{align}
where we have introduced the handy integrals
\begin{align}
\mathcal{J}^{(\sinh)}_{n,m}(T)
&\equiv \int_{0}^{\infty}dp\int_{-1}^{1}d\mu\int_{1}^{\infty}dk\;
\frac{\tfrac12\,(p\mu)^n\,k^{m}}{\sinh^{2}\left(\frac{pk}{4T}\right)},\\
\mathcal{J}^{(\cosh)}_{n,m}(T)
&\equiv \int_{0}^{\infty}dp\int_{-1}^{1}d\mu\int_{1}^{\infty}dk\;
\frac{\tfrac12\,(p\mu)^n\,k^{m}}{\cosh^{2}\left(\frac{pk}{4T}\right)}.
\end{align}

For odd $n$ they vanish by parity.
For even $n$ the $\mu$-integral gives
\begin{equation}
    \int_{-1}^{1}d\mu\,\frac12\mu^n=\frac{1}{n+1},
\end{equation}
so
\begin{align}
\mathcal{J}^{(\sinh)}_{n,m}(T)
&=\frac{1}{n+1}\int_{0}^{\infty}dp\,p^n\int_{1}^{\infty}dk\;
\frac{k^m}{\sinh^{2}\left(\frac{pk}{4T}\right)},\\
\mathcal{J}^{(\cosh)}_{n,m}(T)
&=\frac{1}{n+1}\int_{0}^{\infty}dp\,p^n\int_{1}^{\infty}dk\;
\frac{k^m}{\cosh^{2}\left(\frac{pk}{4T}\right)}.
\end{align}
Now set
\begin{equation}
    z\equiv \frac{pk}{4T},
\end{equation}
which gives
\begin{align}
\mathcal{J}^{(\sinh)}_{n,m}(T)
&=
\frac{(4T)^{n+1}}{n+1}
\int_{1}^{\infty}dk\,k^{m-n-1}
\int_{0}^{\infty}dz\,\frac{z^n}{\sinh^2 z},\\
\mathcal{J}^{(\cosh)}_{n,m}(T)
&=
\frac{(4T)^{n+1}}{n+1}
\int_{1}^{\infty}dk\,k^{m-n-1}
\int_{0}^{\infty}dz\,\frac{z^n}{\cosh^2 z}.
\end{align}
Using
\begin{equation}
    \frac{1}{\sinh^2 z}
    =4\sum_{r=1}^\infty r e^{-2rz},
    \qquad
    \frac{1}{\cosh^2 z}
    =4\sum_{r=1}^\infty (-1)^{r+1}r e^{-2rz},
\end{equation}
we obtain
\begin{align}
    \int_0^\infty dz\,\frac{z^n}{\sinh^2 z}
    &=2^{1-n}n!\,\zeta(n),\\
    \int_0^\infty dz\,\frac{z^n}{\cosh^2 z}
    &=2^{1-n}n!\,(1-2^{1-n})\,\zeta(n).
\end{align}
The remaining $k$-integral is
\begin{equation}
    \int_1^\infty dk\,k^{m-n-1}=\frac{1}{n-m},
\end{equation}
valid for $m<n$, which is the case for the three values $m=-2,0,2$ needed here.
Therefore, for even $n\ge4$,
\begin{align}
    \mathcal J^{(\sinh)}_{n,m}(T)
    &=
    \frac{4(2T)^{n+1}n!\,\zeta(n)}{(n+1)(n-m)},
    \label{eq:J-sinh-final}
    \\
    \mathcal J^{(\cosh)}_{n,m}(T)
    &=
    \frac{4(2T)^{n+1}n!\,(1-2^{1-n})\,\zeta(n)}{(n+1)(n-m)}.
    \label{eq:J-cosh-final}
\end{align}

Substituting Eqs.~\eqref{eq:J-sinh-final} and \eqref{eq:J-cosh-final}
into Eq.~\eqref{eq:Mn-v0-J}, we obtain the closed-form result
\begin{equation}
    \boxed{
    \mathfrak M_n(0)
    =
    \frac{3\,g^2 C_F m_D^2\,2^{\,n-1}T^{\,n-2}(n-1)!\,\zeta(n)}
    {\pi^3 (n+1)\left(N_c+\frac{N_cN_X}{2}+T_F N_f\right)}
    \left[
    \frac{N_c\left(N_X n(n+2)+2(n^2-2n+8)\right)}{8(n-2)(n+2)}
    +\frac{(1-2^{1-n})\,T_F N_f}{n-2}
    \right]
    }
\end{equation}
for all even $n\ge4$, while $\mathfrak M_n(0)=0$ for odd $n$.

As expected, this formula does not apply to $n=2$: the factor $(n-2)^{-1}$ signals the infrared sensitivity of the diffusion coefficient, which is precisely why the $n=2$ moment had to be treated separately in the previous appendix.

\section{Matching-scale invariance}\label{app:invariance}
Choosing an arbitrary matching scale $\Lambda\sim g^0 T$ leads to Eq.~\eqref{eq:SFO-main} changing to
\begin{align}
S_{\rm FO}({\bf L};{\bf v}) = &\frac{g^2 C_F m_D^2}{4\pi}\ln\left(\frac{m_D}{\Lambda}\right)
\Bigg[
\frac{(v^2-1)\arctanh(v)+v}{2v^3}\,
iL_\parallel\left(iL_\parallel+\frac{v}{T}\right)
+\frac{3v^3+(1-v^2)^2\arctanh(v)-v}{4v^3}(iL_\perp)^2
\Bigg]
\nonumber\\
+&\frac{g^2 C_F m_D^2}{4\pi}
\int_0^1 d\mu
\Bigg[
\mu^2\,iL_\parallel\left(iL_\parallel+\frac{v}{T}\right)
+\frac{1-\mu^2}{2}(iL_\perp)^2
\Bigg]\Bigg[
\Phi\left(a_E,b_E\right)
+\frac{v^2(1-\mu^2)}{2(1-v^2\mu^2)}
\Phi\left(a_T,b_T\right)
\Bigg]
\nonumber\\
-&\frac{g^2 C_F}{T} \int_{\bf p}
\left\{ g^2
\frac{e^{i{\bf p}\cdot{\bf L}}-1}{1-e^{-{\bf p}\cdot{\bf v}/T}}\, \mathcal I_{\rm Hard}
-\theta(\Lambda-|{\bf p}|)\, ({\bf p} \cdot i{\bf L}) \left[{\bf p} \cdot\left(i{\bf L} +\frac{{\bf v}}{T} \right)\right] \frac{T}{2\,{\bf p} \cdot{\bf v}} \, \mathcal{I}_{\rm Match}
\right\}.
\label{eq:SFO-lambda}
\end{align}
Differentiating with respect to $\Lambda$ gives
\begin{align}
\frac{\partial}{\partial\Lambda}S_{\rm FO}({\bf L};{\bf v}) = &-\frac{g^2 C_F m_D^2}{4\pi}\frac{1}{\Lambda}
\Bigg[
\frac{(v^2-1)\arctanh(v)+v}{2v^3}\,
iL_\parallel\left(iL_\parallel+\frac{v}{T}\right)
+\frac{3v^3+(1-v^2)^2\arctanh(v)-v}{4v^3}(iL_\perp)^2
\Bigg]
\nonumber\\
&+g^2 C_F \int_{\bf p}
\delta(\Lambda-|{\bf p}|)\,
({\bf p} \cdot i{\bf L}) \left[{\bf p} \cdot\left(i{\bf L} +\frac{{\bf v}}{T} \right)\right]
\frac{\mathcal{I}_{\rm Match}}{2\,{\bf p} \cdot{\bf v}}.
\end{align}
Using
\begin{align}
\frac{\mathcal I_{\rm Match}}{2\,{\bf p}\cdot{\bf v}}
&=
\frac{1}{2pv\mu}
\left[
\frac{2\Gamma_E(\mu v)}{p^4}
+\frac{2v^2(1-\mu^2)\Gamma_T(\mu v)}{p^4(1-v^2\mu^2)^2}
\right]
\nonumber\\
&=
\frac{\pi m_D^2}{2p^5}
\left[
1+\frac{v^2(1-\mu^2)}{2(1-v^2\mu^2)}
\right],
\end{align}
together with
\begin{align}
\int_0^{2\pi}\frac{d\phi}{2\pi}\,
({\bf p}\cdot i{\bf L})\,
\left[{\bf p}\cdot\left(i{\bf L}+\frac{{\bf v}}{T}\right)\right]
=
p^2
\left[
\mu^2\,iL_\parallel\left(iL_\parallel+\frac{v}{T}\right)
+\frac{1-\mu^2}{2}(iL_\perp)^2
\right],
\end{align}
we find
\begin{align}
&\int_{\bf p}
\delta(\Lambda-|{\bf p}|)\,
({\bf p} \cdot i{\bf L})
\left[{\bf p} \cdot\left(i{\bf L} +\frac{{\bf v}}{T} \right)\right]
\frac{\mathcal{I}_{\rm Match}}{2\,{\bf p} \cdot{\bf v}}
\nonumber\\
&\qquad=
\frac{m_D^2}{4\pi\,\Lambda}
\int_0^1 d\mu\,
\left[
\mu^2\,iL_\parallel\left(iL_\parallel+\frac{v}{T}\right)
+\frac{1-\mu^2}{2}(iL_\perp)^2
\right]
\left[
1+\frac{v^2(1-\mu^2)}{2(1-v^2\mu^2)}
\right]
\nonumber\\
&\qquad=
\frac{m_D^2}{4\pi}\frac{1}{\Lambda}
\Bigg[
\frac{(v^2-1)\arctanh(v)+v}{2v^3}\,
iL_\parallel\left(iL_\parallel+\frac{v}{T}\right)
+\frac{3v^3+(1-v^2)^2\arctanh(v)-v}{4v^3}(iL_\perp)^2
\Bigg].
\end{align}
Therefore
\begin{equation}
\frac{\partial}{\partial\Lambda}S_{\rm FO}({\bf L};{\bf v})=0,
\end{equation}
proving matching-scale independence.
\end{document}